\documentclass[journal]{IEEEtran}
\usepackage{graphicx}

\ifCLASSINFOpdf
  
\else
 
\fi

\begin{document}

\title{Complementary Quantum Logic Family Using Josephson Junctions and Quantum Phase-Slip Junctions}

\author{Uday S. Goteti,~\IEEEmembership{Student} Member,~IEEE
Michael C. Hamilton,~\IEEEmembership{Senior Member,~IEEE}

\thanks{Uday S. Goteti and Michael C. Hamilton are with the Department
of Electrical and Computer Engineering, Auburn University, Auburn,
AL, 36830 USA e-mail: mch0021@auburn.edu}

\thanks{Manuscript received October 30, 2018}}

\maketitle

\begin{abstract}
 In this article, we explore a new set of circuits, that incorporate both single-flux-quantum and quantized charge-based complementary quantum logic circuits. Circuits that convert single-flux quantum voltage pulses to quantized charge pulses and vice versa are introduced that lead to circuits that simplify logic and integration operations of individual flux and charge based logic circuits. These include fan-out circuits that enable single flux input to several charge outputs and control gate circuit with charge input controlling flux output. The operation of these circuits is demonstrated in simulations using WRSPICE. An XOR gate implementation is presented as an example to illustrate the operation of these circuits. The developed complementary quantum logic circuits show promise for higher power efficiency and simpler design in the form of fewer junctions for a given logic implementation, leading to the possibility of higher integration density.
\end{abstract}


\IEEEpeerreviewmaketitle

\section{Introduction}

\IEEEPARstart{D}{igital} computing based on superconducting circuits is re-gaining interest for high-performance and energy efficient computation due to potential for high clock rates and low energy operation \cite{manheimer2015cryogenic,holmes2013energy,tolpygo2016superconductor} as concerns about scaling to exa-scale computing grow with traditional CMOS based electronic circuits \cite{exascale,kogge2013exascale}. These superconducting circuits are predominantly based on Josephson junctions (JJs) in the form of single-flux-quantum (SFQ) logic and related variants. Several challenges have been observed in SFQ circuits and attempts have been made to overcome challenges such as increasing power efficiency, reducing static power dissipation, accumulation of jitter, etc. \cite{herr2011ultra,mukhanov2011energy}. Nonetheless, issues corresponding to scalability still exist with SFQ based technologies that require large area cells to accommodate magnetic flux \cite{tolpygo2016superconductor}.

Quantum phase-slip phenomenon has been identified as an exact dual to Josephson tunneling based on charge-flux duality \cite{mooij1}. Several applications of quantum phase-slip junctions (QPSJs) were proposed based on this duality, such as in qubits \cite{mooij2,cqps_nature}, current standards \cite{webster} and other similar applications \cite{graaf2018charge,hriscu1,hriscu_dissertation}. We have explored the use of QPSJs in circuits by establishing their operation in producing a quantized charge of $2e$ in the form of a current pulse with constant area during a single switching event, akin to single-flux quantum generation in a JJ \cite{goteti2018charge,cheng2018spiking}. This operation was used to develop digital logic circuits \cite{goteti2018charge,hamilton2018superconducting} and neuron circuits \cite{cheng2018spiking,cheng2}, which may have potential advantages over JJ-based circuits in realizing higher density circuits, and with possibility of higher power efficiency, albeit with multiple challenges that must be addressed. These challenges correspond to practical realization of controlled quantum phase-slip effects in superconducting nanowires. The effects of phase-slips are susceptible to charge and thermal noise and may require lower temperature operation compared to JJ-based circuits.

In this paper, we introduce complementary quantum logic (CQL) circuits that accommodate both charge-based logic circuits using QPSJs \cite{goteti2018charge} and flux-based circuits using JJs \cite{likharev}. We demonstrate these circuits through simulations in WRSPICE using a SPICE model developed for QPSJs \cite{uday}. CQL circuits are intended to combine both flux-based and charge-based circuits in providing an alternative way to perform logic operations with superconducting circuits. In section II, the basic building blocks of CQL circuits are introduced that utilize charge island \cite{goteti2018charge,hamilton2018superconducting,cheng2018spiking} together with the SFQ loop \cite{likharev}. Circuits that convert quantized flux from SFQ loop to quantized charge from charge island and vice versa are discussed. The conversion circuits together with an additional QPSJ are used to introduce a simple control circuit. Furthermore, fan-out circuits are introduced to use a flux quantum input to drive multiple quantized charge outputs and to use a flux quantum signal to produce a flux and a charge output. In the section III, an example circuit implementing XOR gate logic that employs several of the basic circuit operations implemented in simulation to demonstrate their application. This is followed by a short discussion of potential advantages and challenges of this technology in section IV and conclusion in section V.

\section{Complementary quantum logic circuits}
Complementary quantum logic circuits comprise of both the SFQ pulses encoded in a superconducting loop formed by two JJs and an inductor, as well as the quantized charge pulses encoded on a charge island formed by two QPSJs and a capacitor, as their basic building blocks. The following circuits employ these blocks and the corresponding signals generated, in achieving various operations that are essential in a digital logic family.

\begin{figure}[t!]
    \centering
\includegraphics[width=0.9\linewidth]{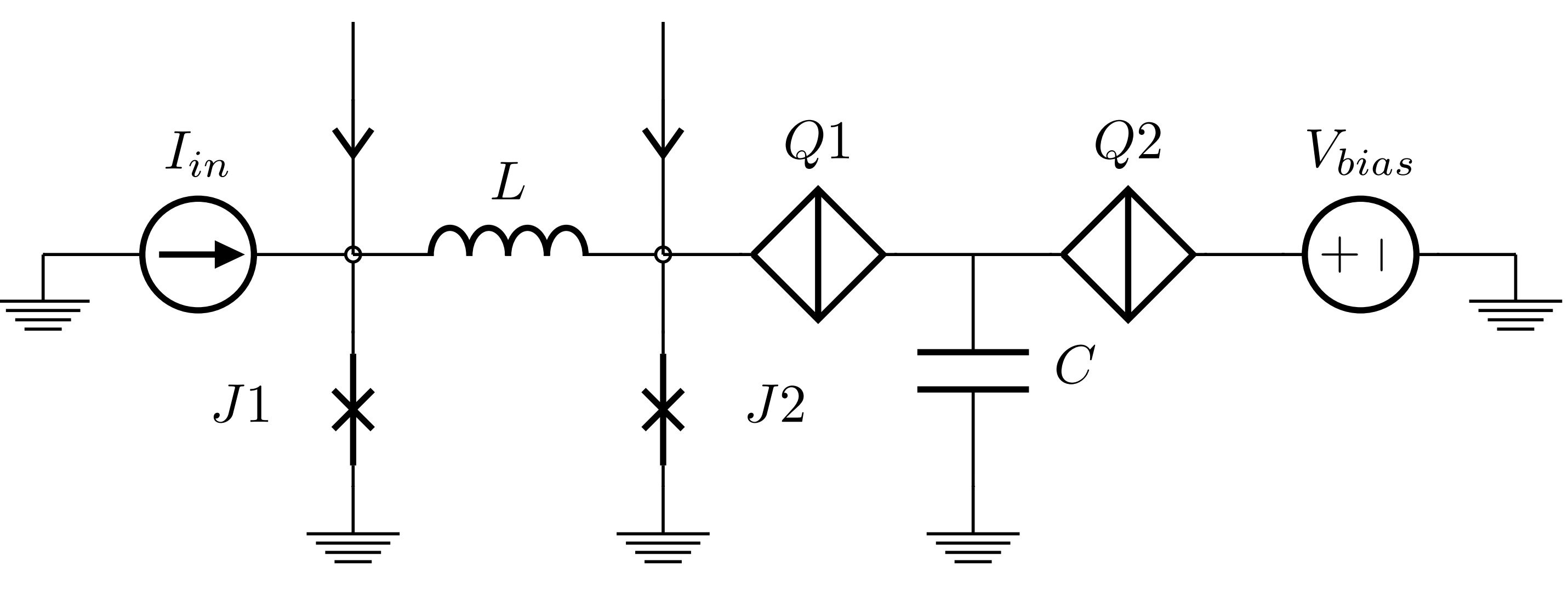}
\caption{SFQ voltage pulse to quantized charge current pulse conversion circuit designed with an SFQ cell and a QPSJ charge island cell. $I_C(J1)=I_C(J2)$, $V_C(Q1)=V_C(Q2)$. DC bias $V_{bias}=1.4V_C$. $I_{bias_{1}}$ = $I_{bias_{2}}$ = 0.7$I_C$.}
\label{phi2e}
\end{figure}

\begin{figure}[t!]
\centering
\includegraphics[width=0.9\linewidth]{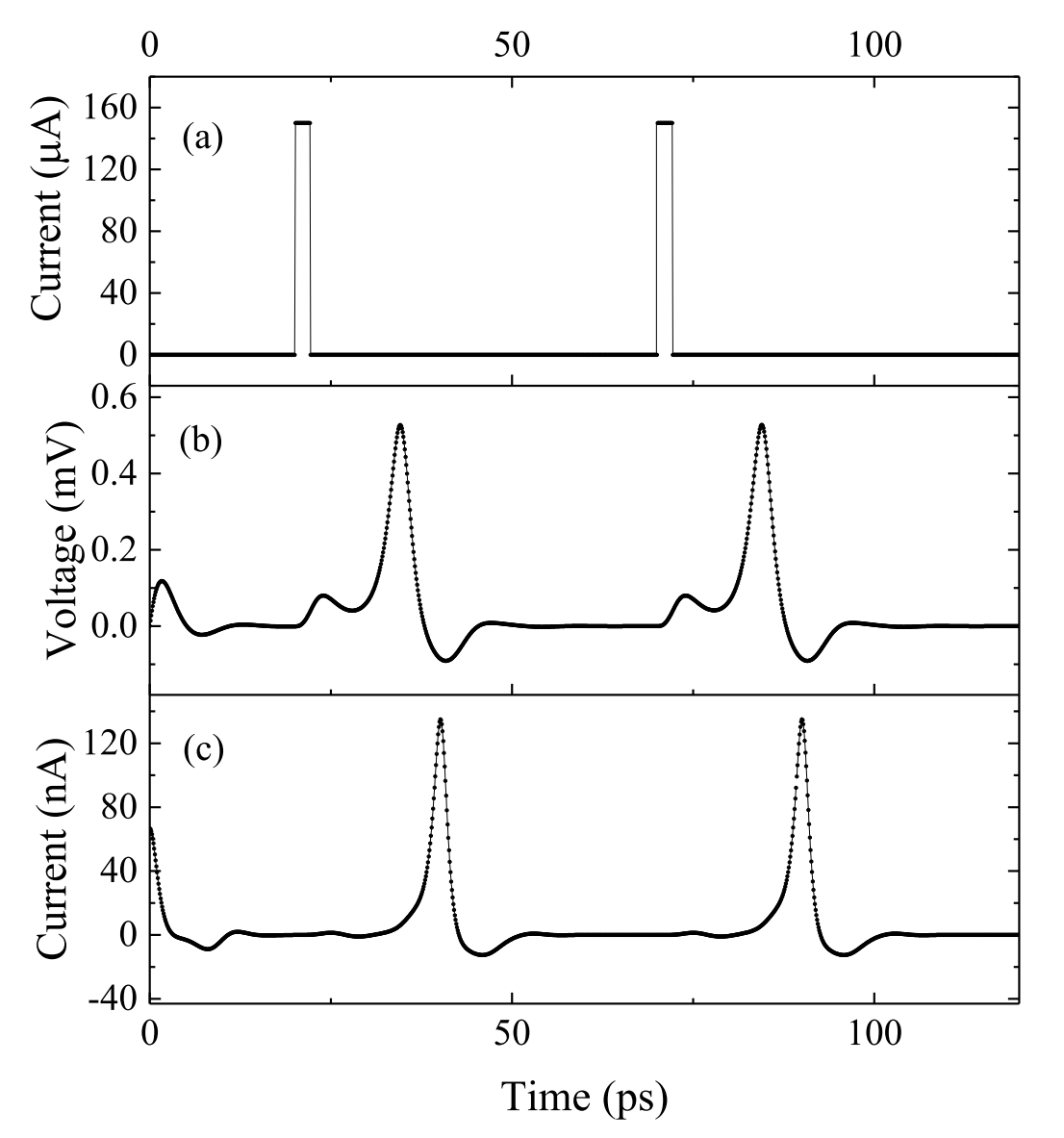}
\caption{Simulation results of flux to charge conversion circuit shown in Fig. \ref{phi2e}. $I_{C}(J1,J2)=100{\mu}$A, $V_{C}(Q1,Q2)=0.7$mV, $L=10.4$pH, $C=0.23$pF. $V_{bias}=1$mV. Magnitude of pulse input $I_{in}=150{\mu}$A. (a) Input current pulses to $J1$ from $I_{in}$. (b) SFQ pulse output from SFQ loop formed by $J1$, $J2$ and $L$ measured at node $1$ of Fig. \ref{phi2e}. (c) Quantized charge output from the charge island formed by $Q1$, $Q2$ and $C$ measured at node $2$ of Fig. \ref{phi2e}.}
\label{flux2charge}
\end{figure}

\subsection{Flux to charge conversion circuit}
The cells corresponding to SFQ loop and charge island can be used in a single circuit to realize flux to charge conversion. The resulting circuit is shown in Fig. \ref{phi2e}. The two identical JJs in the circuit are biased with currents that are $70\%$ of their critical currents $I_C$ and the two identical QPSJs are biased using a DC source $V_{bias}$ with a value of $1.4$ x critical voltage $V_C$ of the junction. An input pulse current drives junction $J1$ to its resistive state generating a voltage pulse corresponding to a flux quantum in the loop formed by $J1$, $L$ and $J2$, that subsequently switches $J2$. The critical voltage of the QPSJs are chosen such that the voltage pulse corresponding to flux quantum at $J2$ can sufficiently drive the QPSJ from its Coulomb blockade state to the conducting state, thereby generating a current pulse of constant area $2e$ at the output. Simulation results of this circuit showing input current pulse, voltage from SFQ loop and the output current pulse from $Q2$ are shown in Fig. \ref{flux2charge}.

\begin{figure}[t!]
    \centering
\includegraphics[width=0.9\linewidth]{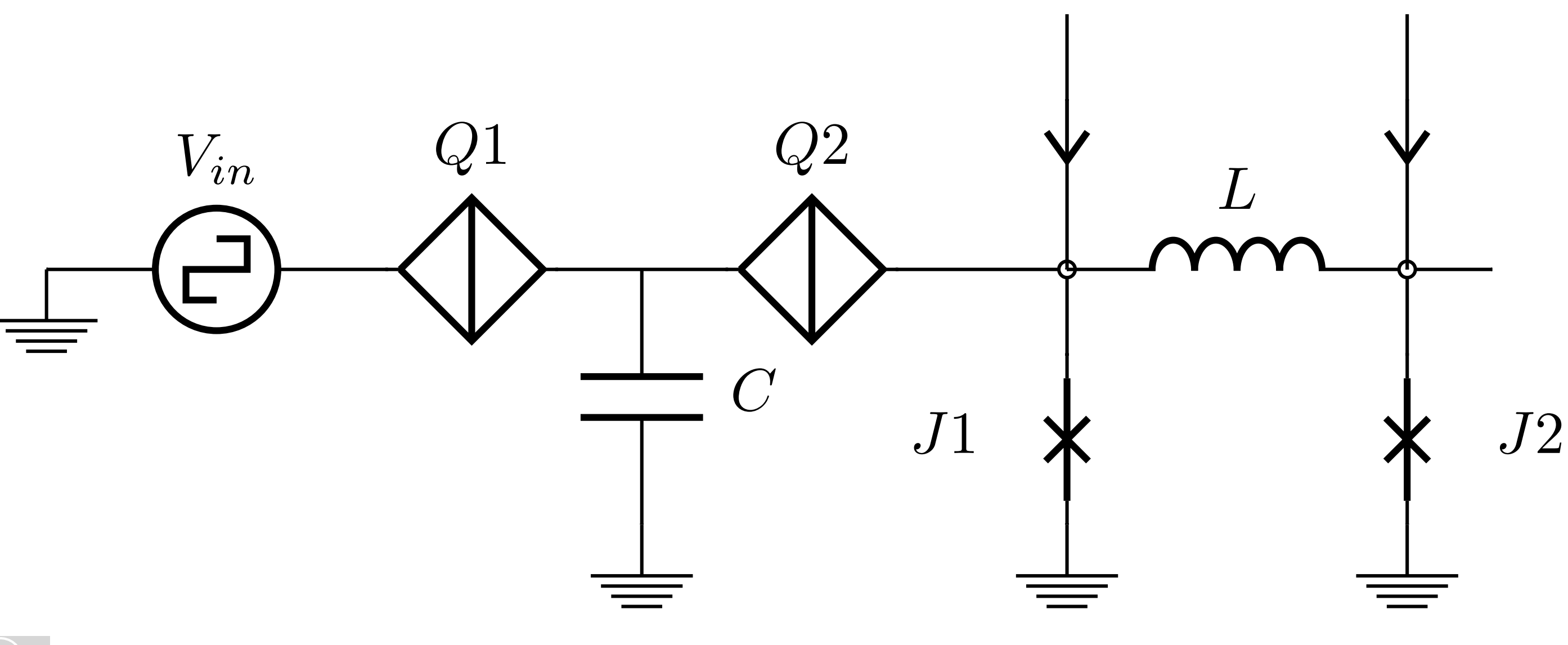}
\caption{Quantized charge current pulse to SFQ voltage pulse conversion circuit designed with an SFQ cell and a QPSJ charge island cell. $I_C(J1)=I_C(J2)$, $V_C(Q1)=V_C(Q2)$. $I_{bias_{1}}$ = $I_{bias_{2}}$ = 0.7$I_C$.}
\label{2ePhi}
\end{figure}

\begin{figure}[t!]
\centering
\includegraphics[width=0.9\linewidth]{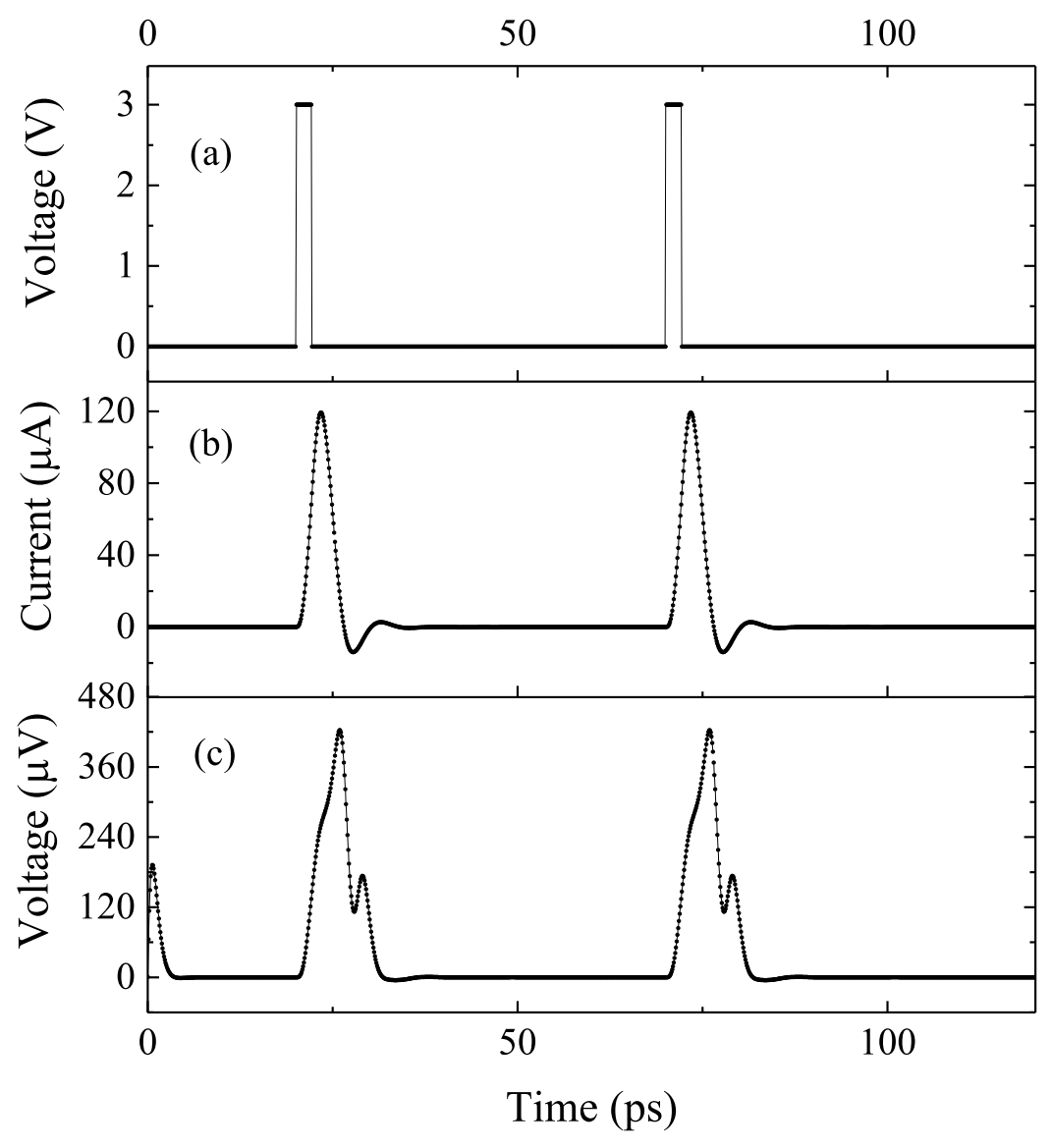}
\caption{Simulation results of charge to flux conversion circuit shown in Fig. \ref{2ePhi}. $I_{C}(J1,J2)=100{\mu}$A, $V_{C}(Q1,Q2)=0.7$mV, $L=10.4$pH, $C=0.23$pF. Magnitude of pulse input $V_{in}=2.8$V. (a) Voltage pulse input to $Q1$ with high voltage amplitude from $V_{in}$. (b) Current pulse output from charge island formed by $Q1$, $Q2$ and $C$ measured at node 1 of Fig. \ref{2ePhi}. (c) Flux output from the SFQ loop formed by $J1$, $J2$ and $L$ measured at node 2 of Fig. \ref{2ePhi}.}
\label{charge2flux}
\end{figure}

\begin{figure}[t!]
    \centering
\includegraphics[width=0.9\linewidth]{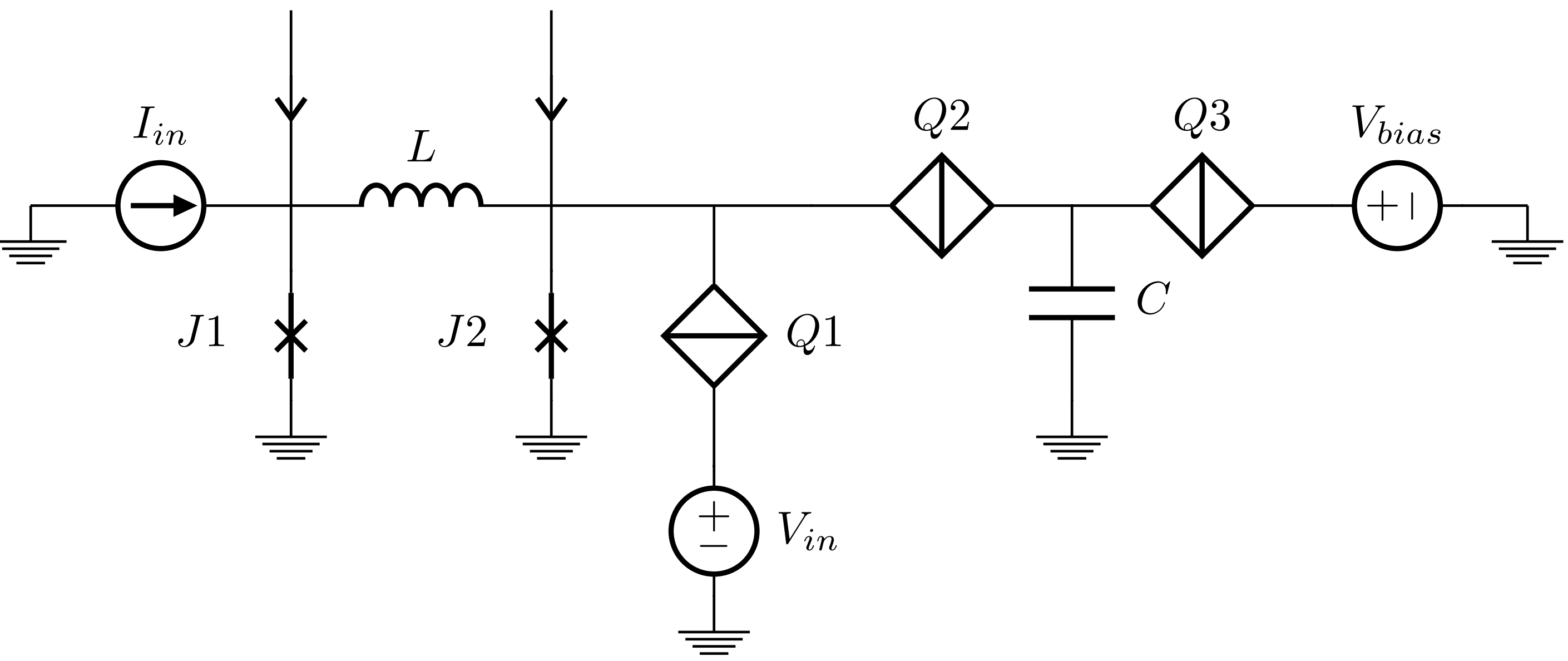}
\caption{Control circuit formed by including an additional QPSJ $Q1$ at the output of SFQ loop with a pulse voltage input. $I_C(J1)=I_C(J2)$, $V_C(Q2)=V_C(Q3)>V_C(Q1)$. DC bias $V_{bias}=1.4V_C$. $I_{bias_{1}}$ = $I_{bias_{2}}$ = 0.7$I_C$.}
\label{inverter}
\end{figure}

\begin{figure}[t!]
\centering
\includegraphics[width=0.9\linewidth]{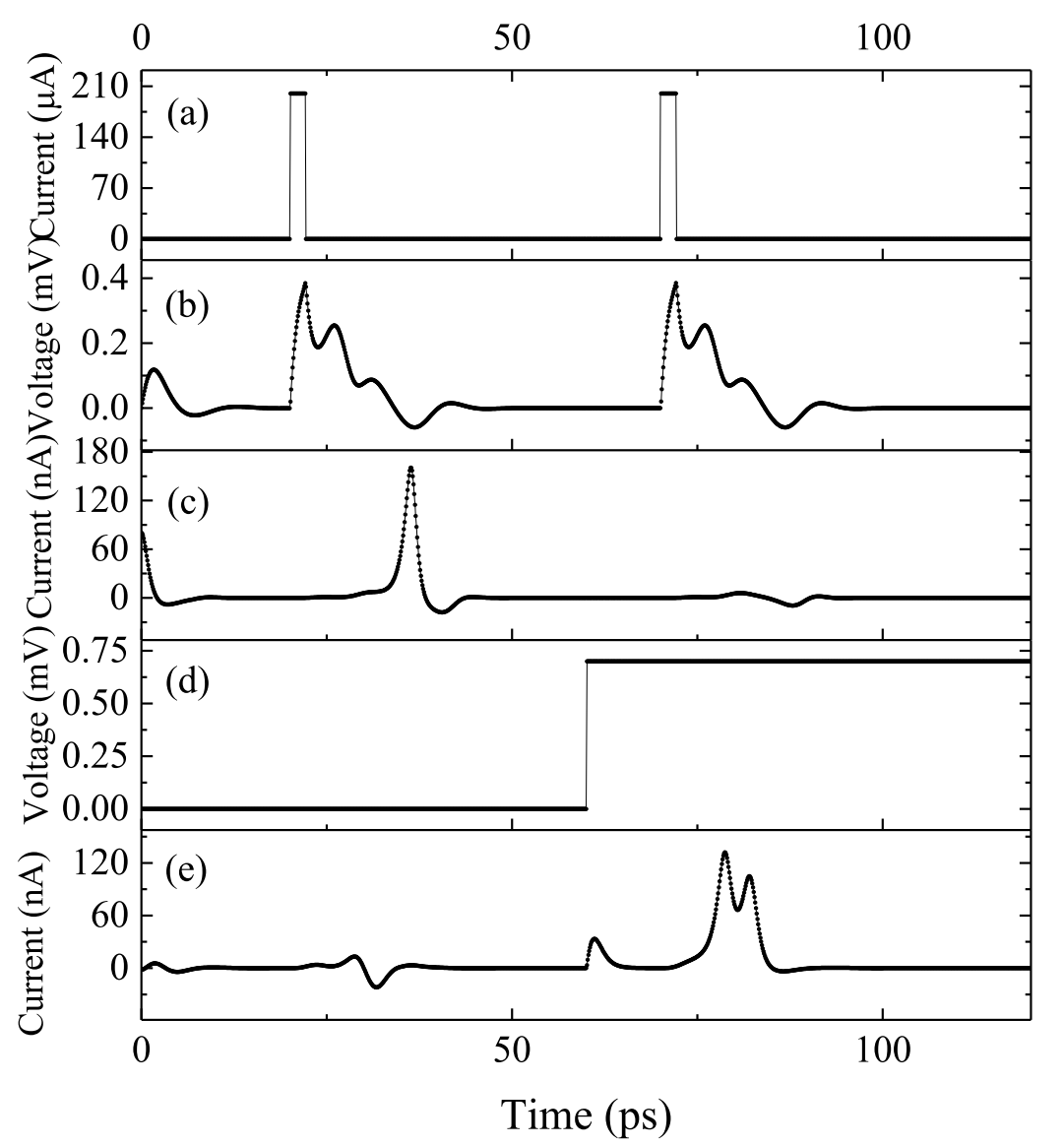}
\caption{Simulation results of the control circuit shown in Fig. \ref{inverter}. $I_{C}(J1,J2)=100{\mu}$A, $V_{C}(Q2,Q3)=1$mV, $V_C(Q1)=0.5$mV, $L=10.4$pH, $C=0.23$pF. $V_{bias}=1$mV. Magnitude of pulse input $I_{in}=200$nA. (a) Current input to the SFQ loop formed by $J1$, $J2$ and $L$ from $I_{in}$. (b) SFQ voltage pulse output from the loop formed by $J1$, $J2$ and $L$ measured at node 1 of Fig. \ref{inverter}. (c) Output at the charge island formed by $Q2$, $Q3$ and $C$ measured at node 3 of Fig. \ref{inverter}. (d) Voltage input $V_{in}$ that controls the output current. (e) Current pulse through the QPSJ $Q1$ when $V_{in}$ is high measured at node 2 of Fig. \ref{inverter}.}
\label{invert}
\end{figure}

\subsection{Charge to flux conversion circuit}
The reciprocal circuit of flux to charge conversion circuit shown in Fig. \ref{phi2e} can be used for charge to flux conversion. The circuit schematic that can achieve such operation is shown in Fig. \ref{2ePhi}, with the parameters identical to the circuit in Fig. \ref{phi2e}. The simulation results are shown in Fig. \ref{charge2flux}. We note that the shape of the SFQ pulse output presented in Fig. \ref{charge2flux}(c) is different from the shape of the SFQ pulse observed in Fig. \ref{flux2charge}(b), but with equal areas (under the curve), each corresponding to single flux quantum. The difference in shape occurs due to the differences in the current pulses (magnitude and duration) applied to the junctions $J1$ in each circuit. This also explains the different SFQ pulse shapes that will be shown in the various other circuits discussed in this paper. Furthermore, the current pulse output from the charge island corresponding to quantized charge of $2e$ is not sufficient to switch large JJs of critical current of 100$\mu$A. Therefore, an input pulse of higher voltage amplitude is used to generate a charge pulse corresponding to $\sim$1000 Cooper pairs, which is sufficient to induce an SFQ pulse at the output for circuit components with the specified parameters. Preliminary simulation results show that, in order to generate a single SFQ pulse output with only a single Cooper pair pulse input, a JJ with a considerably smaller critical current (i.e. up to a few micro-amperes) and a QPSJ with a larger critical voltage (i.e. several hundred milli-volts) are necessary. Practical realization of similar circuits may be challenging with existing technologies, but may be possible with the development of suitable devices or circuits (i.e., with QPSJ-based current amplification). We note that circuits such as these can assist with moving information forward in digital circuits.

\begin{figure}[t!]
    \centering
\includegraphics[width=0.9\linewidth]{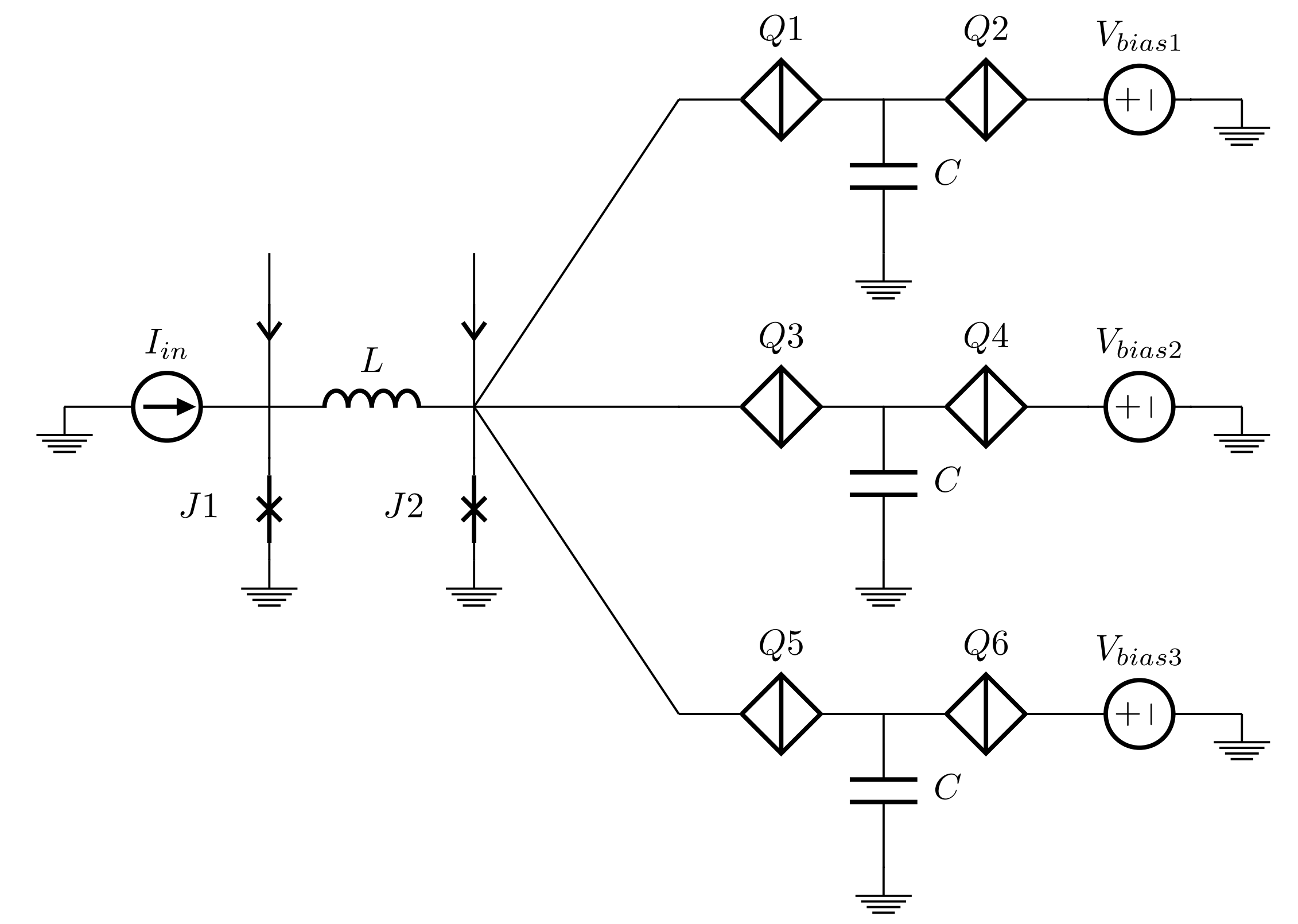}
\caption{A single SFQ input to three quantized charge outputs with loop and island circuit parameters identical to that of Fig. \ref{phi2e}. $I_C(J1)=I_C(J2)$, $V_C(Q1)=V_C(Q2)=V_C(Q3)=V_C(Q4)=V_C(Q5)=V_C(Q6)$. DC bias $V_{bias1}=V_{bias2}=V_{bias3}=1.4V_C$. $I_{bias_{1}}$ = $I_{bias_{2}}$ = 0.7$I_C$.}
\label{jj2qps}
\end{figure}

\begin{figure}[t!]
\centering
\includegraphics[width=0.9\linewidth]{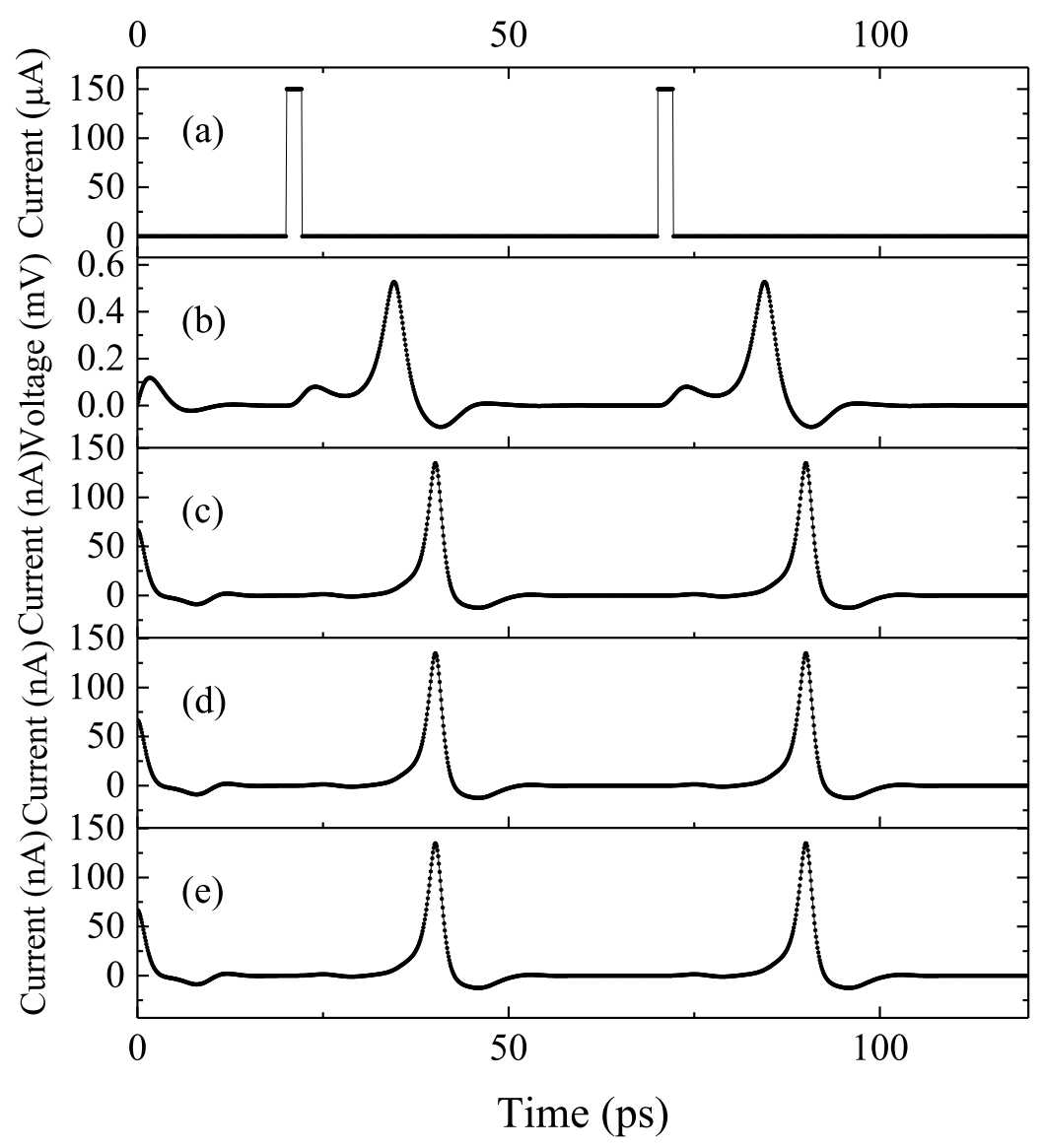}
\caption{Simulation results of fan-out circuit shown in Fig. \ref{jj2qps}. $I_{C}(J1,J2)=100{\mu}$A, $V_{C}(Q1,Q2,Q3,Q4,Q5,Q6)=1$mV, $L=10.4$pH, $C=0.23$pF. $V_{bias1}=V_{bias2}=V_{bias3}=0.7$mV. (a) Current input to the SFQ loop formed by $J1$, $J2$ and $L$ from $I_{in}$. (b) SFQ voltage pulse output from the loop formed by $J1$, $J2$ and $L$ measured at node 1 of Fig. \ref{jj2qps}. (c) Output at the charge island formed by $Q1$, $Q2$ and $C$ measured at node 2 of Fig. \ref{inverter}. (d) Output at the charge island formed by $Q3$, $Q4$ and $C$ measured at node 3 of Fig. \ref{inverter}. (e) Output at the charge island formed by $Q5$, $Q6$ and $C$ measured at node 4 of Fig. \ref{inverter}.}
\label{jj2qpss}
\end{figure}

\begin{figure}[t!]
    \centering
\includegraphics[width=0.9\linewidth]{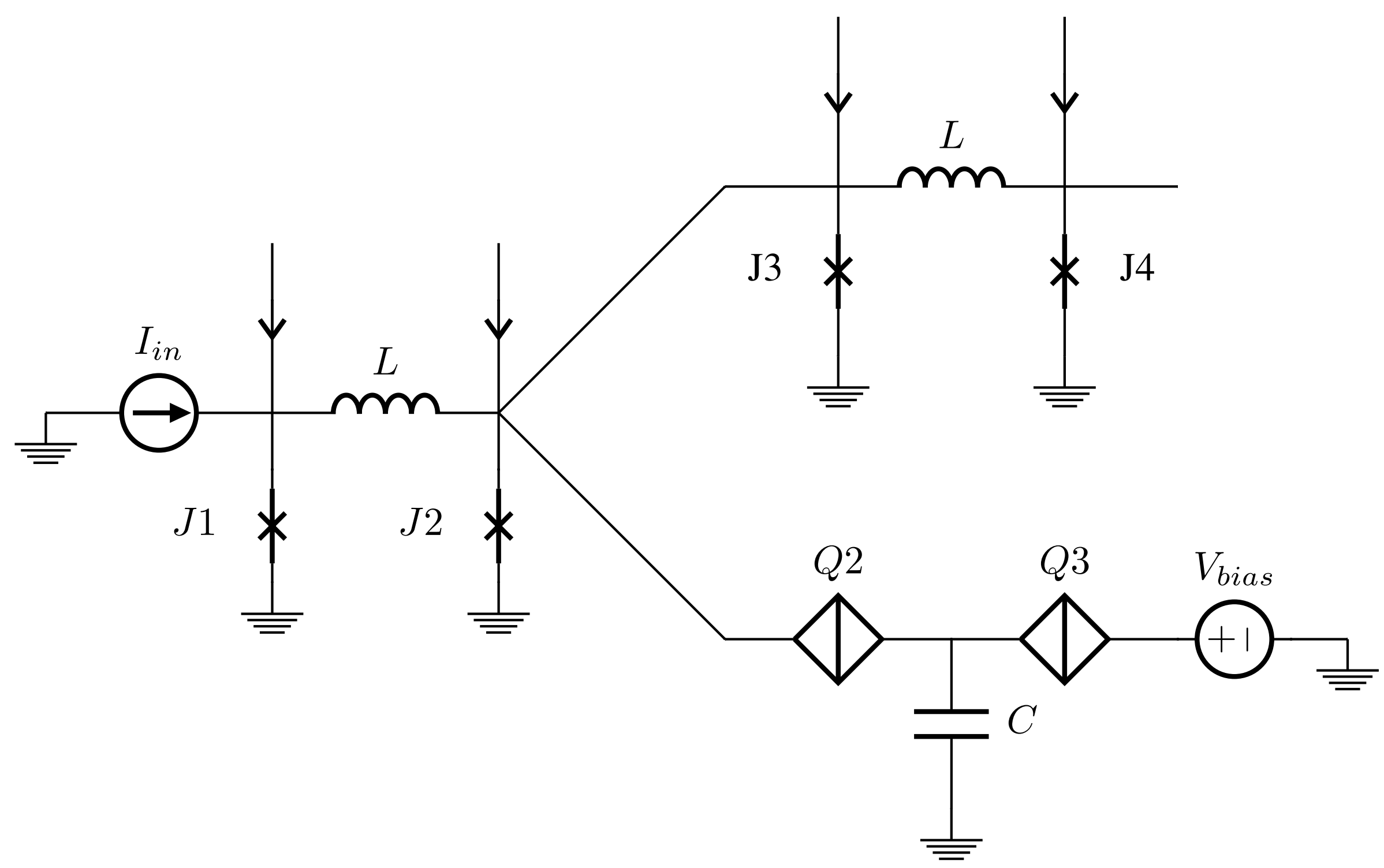}
\caption{A fan-out circuit with single SFQ pulse input with two outputs corresponding to SFQ pulse and quantized charge pulse respectively. $I_C(J1)=I_C(J2)=I_C(J3)=I_C(J4)$, $V_C(Q1)=V_C(Q2)$. DC bias $V_{bias1}=1.4V_C$. $I_{bias_{1}}$ = $I_{bias_{2}}$ = $I_{bias_{3}}$ = $I_{bias_{4}}$ = 0.7$I_C$.}
\label{sfq2sq}
\end{figure}

\begin{figure}[t!]
\centering
\includegraphics[width=0.9\linewidth]{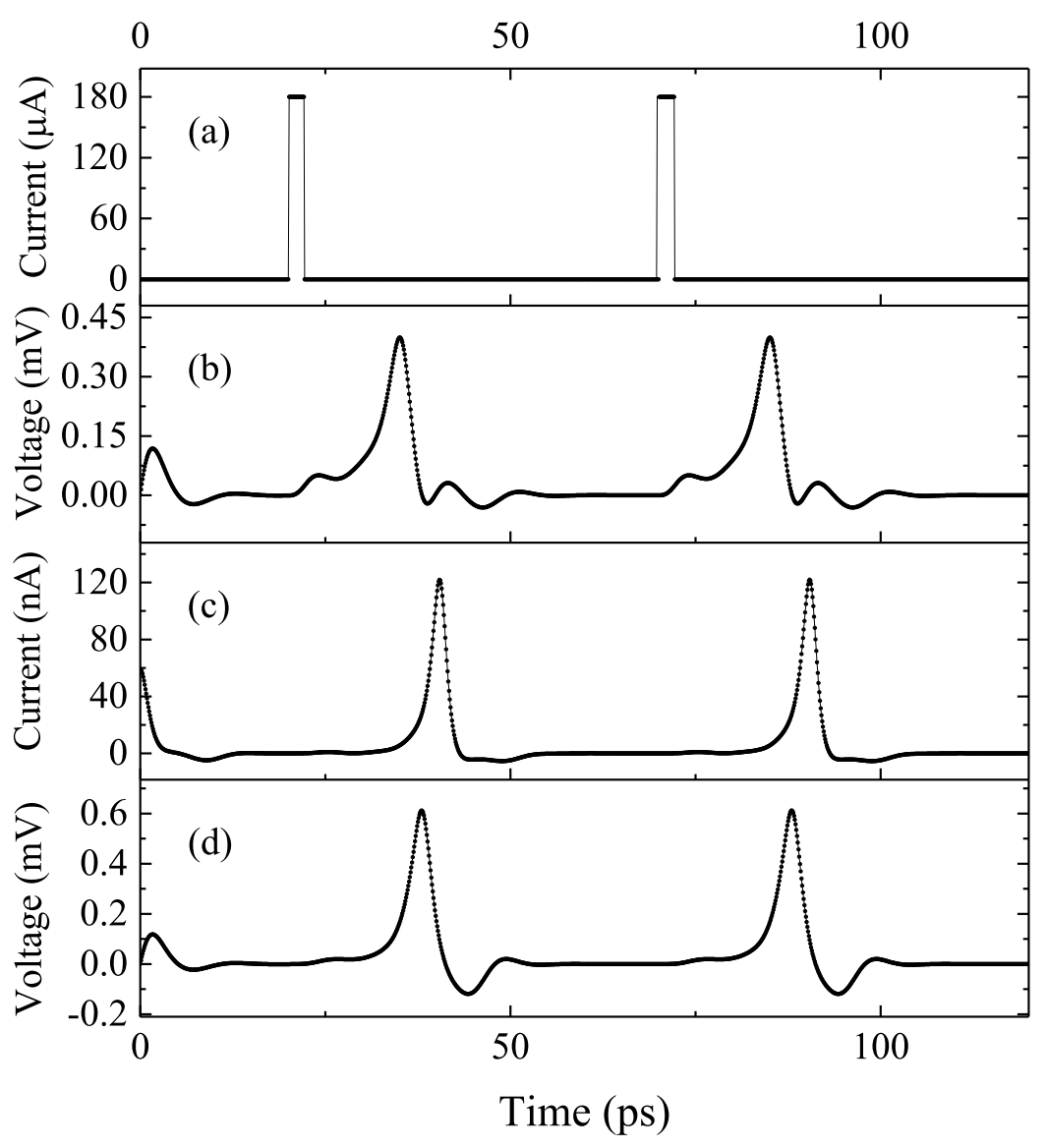}
\caption{Simulation results of fan-out circuit shown in Fig. \ref{sfq2sq}. $I_{C}(J1,J2,J3,J4)=100{\mu}$A, $V_{C}(Q1,Q2)=0.7$mV, $L=10.4$pH, $C=0.23$pF. $V_{bias1}=1$mV. (a) Current input to the SFQ loop formed by $J1$, $J2$ and $L$ from $I_{in}$. (b) SFQ voltage pulse output from the loop formed by $J1$, $J2$ and $L$ measured at node 1 of Fig. \ref{sfq2sq}. (c) Output at the charge island formed by $Q1$, $Q2$ and $C$ measured at node 2 of Fig. \ref{sfq2sq}. (d) Output at the SFQ loop formed by $J3$, $J4$ and $L$ measured at node 3 of Fig. \ref{sfq2sq}.}
\label{sfq2sqs}
\end{figure}

\subsection{Control gate}
The signal flow to the output of the conversion circuits can be controlled using a signal input through a QPSJ similar to control/buffer circuit from \cite{goteti2018charge} resulting in a similar operation. An example control circuit is shown in Fig. \ref{inverter}. The JJs $J1,J2$, QPSJs $Q2,Q3$, along with the inductor $L$ and capacitor $C$ together form the flux to charge conversion circuit shown in Fig. \ref{phi2e}. An additional QPSJ $Q1$ is included along with a step input for switch operation. The critical voltage of $Q1$ is lower than that of $Q2$ and $Q3$, with all other parameters used in the circuit identical to that of Fig. \ref{phi2e}. When the input $V_{in}$ is high, the SFQ pulse from $J2$ switches junction $Q1$ before junction $Q2$, resulting in no current pulse at the output. When the input $V_{in}$ is low, the SFQ pulse from $J2$ switches $Q2$, resulting in flux to charge conversion. Simulation results illustrating this operation are shown in Fig. \ref{invert}. Similar operation can be implemented with charge to flux conversion circuit.

\subsection{Fan-out circuits}
Fan-out circuit schematic is useful to drive several gates with charge/flux input connected to a flux/charge outputs. Conversion from flux to charge and vice versa enables using a single input to drive several outputs without decrease in the pulse amplitudes. Furthermore, it is possible to split the input to either charge or flux output in the same circuit. These two operations are demonstrated in the circuits below.

\subsubsection{SFQ input splitter to multiple quantized charge outputs}
The circuit shown in Fig. \ref{jj2qps} can be used to split an SFQ pulse input to three quantized charge outputs. This operation can be extended to a higher number outputs by including more charge islands at the output of SFQ loop. Furthermore, there are no restrictions on the junction parameters irrespective of the number of outputs when the islands are biased with sufficient voltage. This is because the voltage drop at node 1 of Fig. \ref{jj2qps} due to leakage current through connected charge islands is negligible. The simulation results of the circuit shown in Fig. \ref{jj2qps} are shown in Fig. \ref{jj2qpss}. The reciprocal circuit operation, i.e. from charge input to several flux outputs is possible provided the critical currents of JJs are significantly lower (i.e. on the order of a few micro-amperes). Practical realization may be challenging with present technologies, as mentioned in Section II.B., without internally amplifying the charge input.

\subsubsection{SFQ input splitter to SFQ and charge quantum output splitter}
The circuit shown in Fig. \ref{sfq2sq} can be used to split a single SFQ pulse input to an SFQ pulse output and a quantized charge pulse output. Simulations results illustrating this operation are shown in Fig. \ref{sfq2sqs}. 

\section{Logic circuit example}
\begin{figure}[t!]
    \centering
\includegraphics[width=0.9\linewidth]{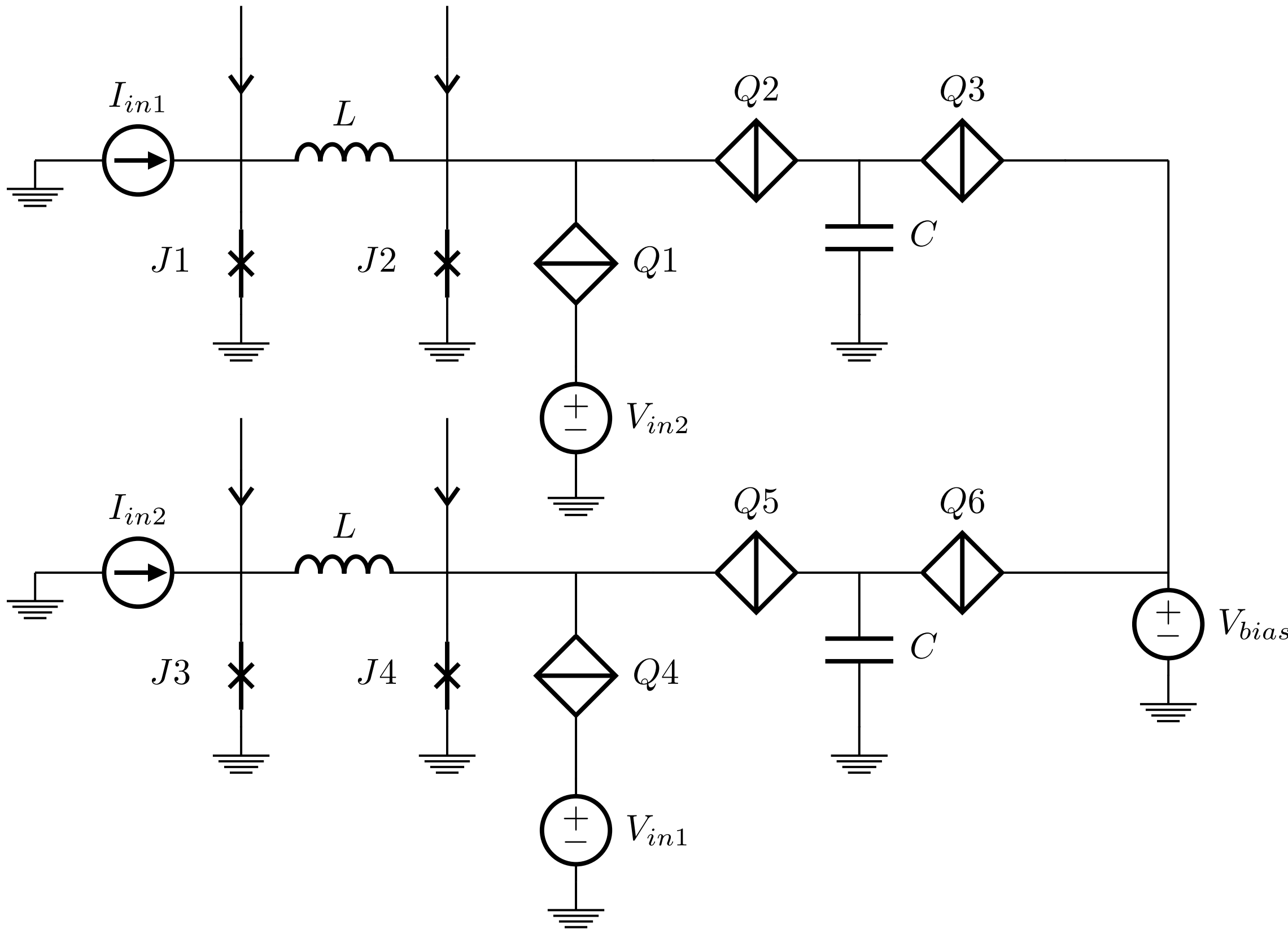}
\caption{XOR gate operation achieved by introducing additional QPSJs $Q1$ and $Q4$ at the output of SFQ loops, with a pulse voltage input.}
\label{xor}
\end{figure}

\begin{figure}[t!]
\centering
\includegraphics[width=0.8\linewidth]{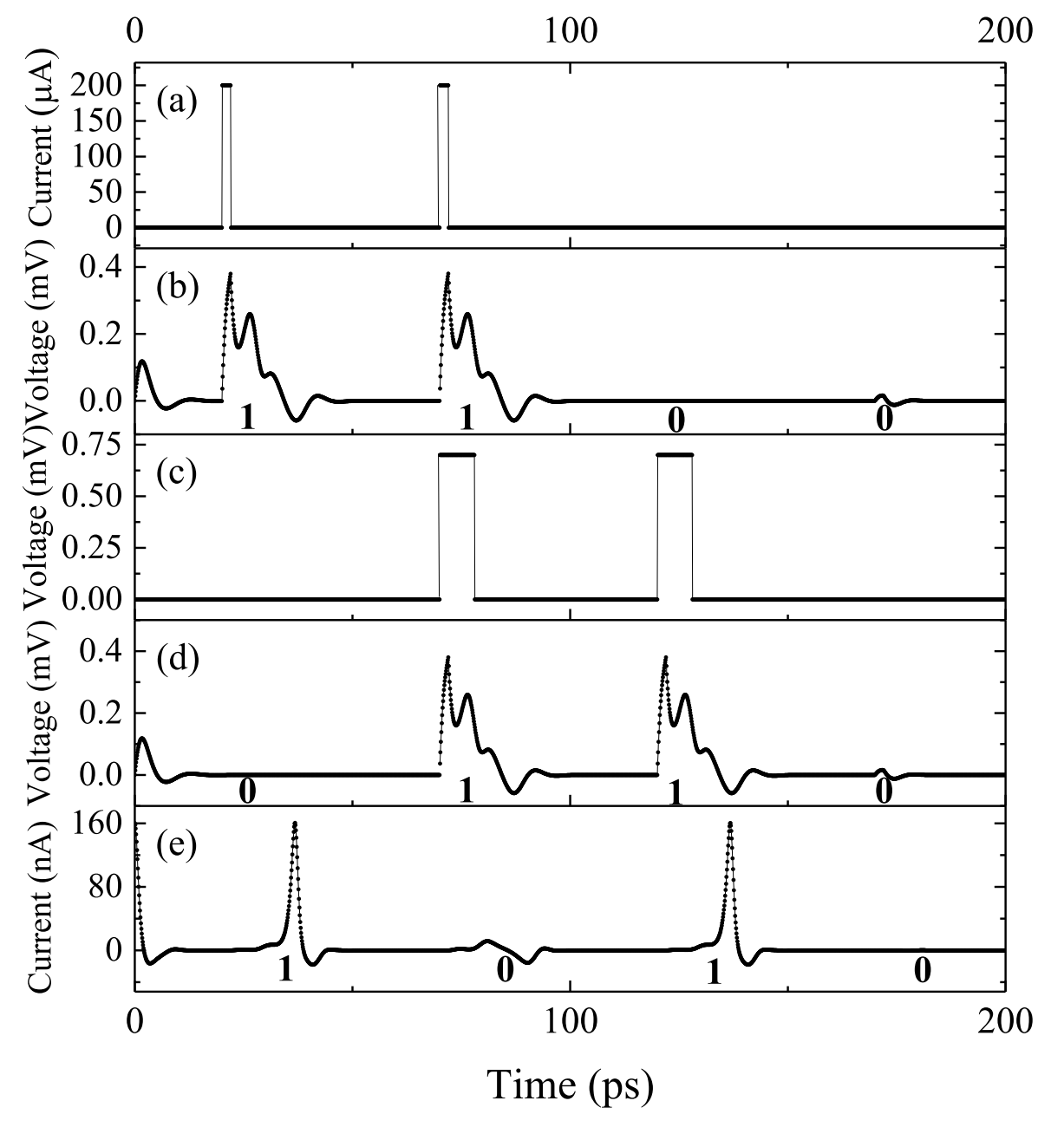}
\caption{Simulation results of XOR gate shown in Fig. \ref{inverter}. (a) Input current pulses $I_{in1}$ to generate SFQ pulses in loop $J1$, $L$ and $J2$. (b) Input 1 of XOR gate from SFQ loop $J1$, $L$ and $J2$ measured at node 1 of Fig. \ref{xor}. (c) Input voltage pulse from $V_{in2}$ to generate quantized charge pulses at $Q1$. (d) Input 2 of XOR gate  from SFQ loop $J3$, $L$ and $J4$ measured at node 2 of Fig. \ref{xor}. (e) Output of XOR gate as quantized charge current pulses from $Q3$ and $Q6$ measured at node 3 of Fig. \ref{xor}.}
\label{xors}
\end{figure}

A two input XOR gate can be implemented using two flux to charge conversion circuits combined with control gates in parallel. Four inputs are applied to the SFQ cells at junctions $J1$ and $J3$, and at the junctions $Q1$ and $Q4$. The input 1 at junction $J1$ and the input at $Q4$ are high at the same time, and the input 2 at $J3$ and the input at $Q1$ are high at the same time. This is illustrated in the circuit shown in Fig. \ref{xor}, and the corresponding simulation results are shown in Fig. \ref{xors}. During practical implementation, same input signals can be used in these cases with appropriate charge/flux conversion circuits. When both the inputs are '1', QPSJs $Q1$ and $Q4$ are switched, therefore the signals generated in both SFQ cells do not travel into the QPSJ charge islands. This results in the output '0'. When only one of the inputs is '1', the SFQ pulse generated in the JJ corresponding to that input is converted to quantized charge at the corresponding island, generating the output '1'. The output is '0' when both the inputs are '0', as none of the junctions are switched, resulting in no SFQ pulses. 

\section{Discussion}
CQL circuits provide an a new way to perform digital logic operations in superconducting electronics that are predominantly based on JJs alone, by utilizing QPSJs, with some potential advantages. The charge islands formed by QPSJs can generate quantized charge pulses that are similar to SFQ pulses, but the switching energy of QPSJs to generate a current pulse is estimated to be order of 1-5 zJ. This is considerably smaller than that of currently available JJ technologies which dissipate energy in the order of several aJ. Using JJs and QPSJs together enables convenient fan-out to multiple outputs without a loss of output signal amplitude, in addition to requiring fewer junctions to implement a logic operation compared to JJ-based circuits. 

Although CQL circuits may provide significant advantages, challenges exist in practical realization of these circuits, pertaining to the material and design of QPSJs for controlled generation of quantum-phase slips in nanowires, along with a possible need for lower operating temperatures (perhaps below 1 K). Other potential issues exist such as interference of charge noise with the charge on islands. The extent of these issues and possible solutions may only be evident after sufficient investigation through experiments.

\section{Conclusion}
A new family of circuits is introduced that combines the SFQ operation of JJs and quantized charge operation of QPSJ based circuits to perform digital logic. These circuits provide an alternative way to perform logic operations that may significantly simplify the design when compared to JJ-based logic families, therefore may improve flexibility when these circuits are scaled to peta and exa-scale computers. Flux to charge conversion circuits and vice versa are presented that can be interfacing circuits between JJ and QPSJ based logic circuits. Logic operations such as an inverter and fan-out to multiple outputs are demonstrated as examples to illustrate the applications of these logic circuits. However, substantial developments in technology are required for physical realization of single QPSJs that exhibit these properties, as well as in testing the circuits discussed in this paper.

\ifCLASSOPTIONcaptionsoff
  \newpage
\fi

\bibliographystyle{plain}

\end{document}